\documentclass{article}
\usepackage{amsfonts}
\usepackage{amsmath}
\usepackage{geometry}
\usepackage{amssymb}
\usepackage{graphicx}

\providecommand{\U}[1]{\protect\rule{.1in}{.1in}}
\providecommand{\U}[1]{\protect\rule{.1in}{.1in}}
\providecommand{\U}[1]{\protect\rule{.1in}{.1in}}

\input{tcilatex}
\begin{document}

\begin{center}
\bigskip

{\LARGE \ {\large \textbf{Oscillator Algebra of Chiral Oscillator}}\\[1cm]
}

\bigskip

H. G\"{u}mral{\LARGE \ }

Department of Mathematics, Yeditepe University

34755 Ata\c{s}ehir, Istanbul, Turkey,

\bigskip hgumral@yeditepe.edu.tr
\end{center}

\bigskip

\textbf{Abstract: }For the chiral oscillator described by a second order and
degenerate Lagrangian with special Euclidean group of symmetries, we show,
by cotangent bundle Hamiltonian reduction, that reduced equations are
Lie-Poisson on dual of oscillator algebra, the central extension of special
Euclidean algebra in two dimensions. This extension, defined by symplectic
cocycle of special Euclidean algebra, seems to be an enforcement of
reduction itself rooted to Casimir function.

\smallskip



\bigskip

\bigskip

\section{Introduction}

In this work, we shall show that the Euler-Lagrange equations for the chiral
oscillator from the second order and degenerate Lagrangian \cite{luk}-\cite%
{hg22} 
\begin{equation}
L_{Chiral}\left[ x,y\right] =-\frac{\lambda }{2}(\dot{x}\ddot{y}-\dot{y}%
\ddot{x})+\frac{m}{2}(\dot{x}^{2}+\dot{y}^{2})  \label{chiral}
\end{equation}%
can be described as a Hamiltonian vector field tangent to a curve in certain
codimension four submanifold of eight dimensional phase space $T^{\ast }TM.$
The relevant Poisson structure is that of dual $\mathfrak{osc}^{\ast }%
\mathfrak{=(se(2)\times 
\mathbb{R}
})^{\ast }$ of oscillator algebra, which is the central extension of algebra 
$\mathfrak{se(2)}$ of special Euclidean group of symmetries of configuration
space. This structure is obtained by Poisson reduction from
Ostrogradskii-Hamiltonian structure \cite{ostro}-\cite{crampin86} of $%
T^{\ast }TM$ constructed by means of Dirac-Bergman theory \cite{dirac}-\cite%
{sal}. Moreover, on three dimensional level sets of Casimir of $\mathfrak{%
se(2)}^{\ast }$, namely, lenght of linear momentum vector, solutions are
intersections of reduced Hamiltonian function and Casimir function of
reduced bracket.

The chiral Lagrangian (\ref{chiral}) is invariant under translations in $x$
and $y$ directions and rotations on their plane $M$. Together they form the
group $SE(2)$ of Euclidean motions. Lifted actions have momentum maps with
components $(\mu ,\mathbf{p}^{0})$, angular and linear momenta,
respectively. In these coordinates, Lie-Poisson structure on $\mathfrak{se(2)%
}^{\ast }$ admits the Casimir function $\mid \mathbf{p}^{0}\mid ^{2}=:l^{2}$%
. The reduced submanifold for fixed regular values of $(\mu ,\mathbf{p}%
^{0},l^{2})$ is four dimensional and acquires the constant function $l^{2}$
as one of its coordinates. This endows the reduced submanifold with
Lie-Poisson structure on the dual of the central extension $\mathfrak{%
osc=se(2)\times 
\mathbb{R}
}$. This enforcement of reduction does not change trivial dynamics of $l^{2}$
and we are left with a three dimensional Hamiltonian system whose solution
can be described geometrically.

To achieve these results, we shall apply cotangent bundle Hamiltonian, or
Poisson reduction \cite{mw}-\cite{marle14}. As the chiral oscillator
Lagrangian is second order, we will use Ostrogradskii approach to define
Hamiltonian phase space. Degeneracy, on the other hand, requires the
Dirac-Bergman analysis to obtain Hamiltonian structure, namely, Poisson
brackets and Hamiltonian functions. These are already available in the
literature \cite{luk},\cite{luk99},\cite{hg22}. Thus, the main computations
involve parametrizations of level sets of regular values of momentum map in
such a way that Hamiltonian functions describing dynamics are also functions
of these coordinates that must be invariant under the symmetry group.

Oscillator algebra was first studied by Streater \cite{streater67} in the
framework of representation theory. In reduction theories, general framework
for Lie-Poisson structures of central extension of Lie algebras with cocyles
was explained in Marsden et al. \cite{bigstages07}. In Vankerschaver, Kanso
and Marsden \cite{vkm10}, Lie-Poisson structure on dual of oscillator
algebra was obtained by reduction with $SE(2)$ of cotangent bundle
symplectic structure with magnetic term. It was concluded that the extension
is caused by gyroscopic forces related to magnetic term such as Lorentz and
Kutta-Zhukowski forces. To our knowledge, chiral oscillator is yet another
case in which extention to oscillator algebra follows from reduction, this
time without magnetic term.

\section{Reductions of Lagrangians}

In his 1901 paper \cite{poi01}, Poincar\'{e} considered a first order
Lagrangian on configuration space $N$ with symmetry group $G$ and
successfully replaced second order Euler-Lagrange equations by first order
equations, now called Euler-Poincar\'{e} equations on Lie algebra $\mathfrak{%
g}$. This paper is elegantly described by Marle \cite{marle14},\cite{marle}
in the modern language of symplectic geometry that we will follow for the
present discussion. The gist of Poincar\'{e}'s argument is to have a
trivialization of tangent bundle $TN\approx N\times \mathfrak{g}$ and
restrict the Lagrangian function to fibers. This requires large enough
symmetry group depending on dimension of configuration space.

Given a Lagrangian function\ $L:TN\longrightarrow 
\mathbb{R}
$ with symmetry group $G$, its Lie algebra $\mathfrak{g}$ acts on $N$ by
fundamental vector fields $F:\mathfrak{g}\longrightarrow TN$ and on $TN$ by
tangent lifts of $F_{X}$ for each $X\in \mathfrak{g}$. If dim $G$ \ $\geq $
\ dim $N$, one can replace each fiber of $TN$ by Lie algebra and find a map $%
\alpha :TN\longrightarrow \mathfrak{g}$ to restrict $L=L_{red}\circ \alpha $
to a reduced function $L_{red}$ on fibers. The map $\alpha $, being a Lie
algebra valued one-form, is called a mechanical connection \cite{bigstages07}%
. $L_{red}$ together with induced variational principle on $\mathfrak{g}$
results in Euler-Poincar\'{e} equations \cite{bigstages07},\cite{marle14},%
\cite{marle}.

In \cite{marle}, Marle showed that Euler-Poincar\'{e} equations can
intrinsically be expressed in terms of Legendre $\mathbb{F}%
L:TN\longrightarrow T^{\ast }N$ and momentum $\mathbb{J}:T^{\ast
}N\longrightarrow \mathfrak{g}^{\ast }$ maps because they conveniently
replace fiber or, vertical derivative (with respect to Lie algebra variable)
of Lagrangian function. Proving that the existence of restriction $L_{red}$
is connected with constant values of momentum map on the image of Legendre
map (c.f. Lemma 1 in \cite{marle}), he concluded that Euler-Poincar\'{e} and
Hamilton's equations are equivalent. It has already been stated in \cite%
{marsden} that reductions of Lagrangian formulations are indeed Lagrangian
versions of the Marsden-Weinstein reduction \cite{mw}-\cite{marle14} of the
phase space Hamiltonian formalism. These results and ideas are the main
motivations for the present work where we implement the existing Hamiltonian
structures from \cite{luk}-\cite{cruz},\cite{hg22} for the second order
chiral oscillator Lagrangian to obtain reduced equations.

For the remaining case of dim $G$ \ $<$ \ dim $N$ one starts with the
horizontal-vertical decomposition $TN=HN\oplus VN$ of tangent bundle and
identifies the vertical part with the Lie algebra of symmetries. In general
cases, equations of motion consist of horizontal part defined by usual
Euler-Lagrange equations and vertical part on $\mathfrak{g}$ which,
collectively called Lagrange-Poincar\'{e} equations.

On the Hamiltonian side, we have Hamiltonian function $H:T^{\ast
}N\longrightarrow 
\mathbb{R}
$ which may be obtained from Lagrangian function $L$. Lie algebra $\mathfrak{%
g}$ acts on $T^{\ast }N$ by cotangent lifts which are Hamiltonian vector
fields with respect to canonical Hamiltonian structure of $T^{\ast }N$. Note
that tangent lifts can also be made into Hamiltonian vector fields with
appropriate Hamiltonian structure obtained from given Lagrangian. For large
enough symmetry group, one can replace each fiber of $T^{\ast }N$ by $%
\mathfrak{g}^{\ast }$ and obtain the trivialization\ $T^{\ast }N\approx
N\times \mathfrak{g}^{\ast }$. Restriction of Hamiltonian function $H$ to
fibers $\mathfrak{g}^{\ast }$ is now natural by the momentum map\ $\mathbb{J}%
:T^{\ast }N\longrightarrow \mathfrak{g}^{\ast }$ defined, for each $X\in 
\mathfrak{g}$, as%
\begin{equation*}
<\mathbb{J}(\mathbf{z}),X>=J_{X}(\mathbf{z})\text{, \ \ \ \ \ }%
i_{F_{X}^{lift}(\mathbf{z})}\omega =dJ_{X}(\mathbf{z})
\end{equation*}%
where $\mathbf{z}$ is the coordinates on $T^{\ast }N$, $<\cdot ,\cdot >$ is
the pairing of $\mathfrak{g}$ and $\mathfrak{g}^{\ast }$, the vector field $%
F_{X}^{lift}(\mathbf{z})$ is the appropriate lift of $X\in \mathfrak{g,}$
and $\omega $ is the symplectic form on $T^{\ast }N$. Writing $%
H=H_{red}\circ \mathbb{J}$ one obtains the reduced Hamiltonian function $%
H_{red}$ on $\mathfrak{g}^{\ast }$.

Details and examples of reductions of both Lagrangian and Hamiltonian
systems along with historical development of subjects and very extensive
literature can be found in Marsden, Ratiu and Scheurle \cite{marsden}. Here,
we continue with discussion on our intension to apply Poincar\'{e}'s ideas
for second order Lagrangians aiming at an implementation of them for chiral
oscillator Lagrangian. See for example, \cite{gay10,gay11} for other
geometric treatments of reductions of higher order Lagrangians.

Let $M$ be the configuration manifold for a second order Lagrangian \ $%
L:T^{2}M\longrightarrow 
\mathbb{R}
$ with symmetry group $G$. The second order tangent space $T^{2}M$ is not
even a vector bundle but can only be embedded into iterated tangent bundle $%
TTM.$ This means $G$ acts on $L$ not by tangent lifts but by prolongations
which is useful in deriving Noetherian conservation laws from symmetries 
\cite{olver}. Ostrogradskii formalism carries higher order Lagrangian
dynamics to a geometric setting on a cotangent bundle. In particular, our
implementation of Hamiltonian reduction procedure for second order theories
will start with replacing $N$ above with $TM$. On the canonically symplectic
space $T^{\ast }TM$, symmetry algebra $\mathfrak{g}$ acts by cotangent lifts
of tangent lifts of fundamental vector fields and this action is generated
by Hamiltonian vector fields with Hamiltonian functions forming components $%
\mathbb{J}_{a}$ of momentum map $\mathbb{J}:T^{\ast }TM\longrightarrow 
\mathfrak{g}^{\ast }$. We will then apply reduction in the sense of Marsden
and Weinstein \cite{mw}-\cite{marle14}.

Thus, we start using Ostrogradskii formalism to construct canonical Poisson
bracket $\left\{ ,\right\} _{C}$ and canonical Hamiltonian function $H^{C}$
on $T^{\ast }TM$. For degenerate Lagrangians, we use Dirac-Bergman
constraint analysis to obtain Dirac's total Hamiltonian function $H^{D}$ and
Dirac's bracket $\left\{ ,\right\} _{D}$. The corresponding Hamiltonian
structures on $T^{\ast }TM$ will be $(H^{D},\left\{ ,\right\} _{C})$ and $%
(H^{C},\left\{ ,\right\} _{D})$. These are equivalent ways of writing
Hamilton's equations. The algebraic structure on fibers $\mathfrak{g}^{\ast
} $ is obtained by evaluating the canonical Poisson bracket on $T^{\ast }TM$
for components $\mathbb{J}_{a}$ of momemtum map. This gives Lie-Poisson
structure. Since $\mathbb{J}$ is a Poisson map \cite{mr}, the reduced space $%
\mathbb{J}^{-1}(\nu )$ for regular fixed values of momenta $\nu \in 
\mathfrak{g}^{\ast }$ carries the same agebraic structure. The main
computation for the characterization of $\mathbb{J}^{-1}(\nu )$ involves
choice of parametrization by $G$ invariant functions $J_{X_{a}}$ such that
Hamiltonian functions $H^{D}$ or $H^{C}$ can be expressible in terms of $%
J_{X_{a}}$s. As a matter of fact, this was shown by Marle to be one of the
conditions for reduction to occur and, for Euler-Poincar\'{e} equations to
be equivalent to reduced Hamilton's equations (c.f section 5 of \cite%
{marle14}). To this end, we recall that invariant functions defined as
Hamiltonian functions of usual cotangent lifts are linear in fiber
coordinates of $T^{\ast }TM$. We will see that one needs more general
invariant functions to be used as coordinates on $\mathbb{J}^{-1}(\nu )$.
That means, actions on $T^{\ast }TM$ may be generated by lifts of actions on
configuration manifold more appropriate than tangent and cotangent lifts.

Evaluating the canonical Poisson bracket of these invariant functions we
will obtain reduced Poisson bracket algebra on $\mathbb{J}^{-1}(\nu )$. With
reduced Hamiltonian function and reduced bracket we find reduced equations
of motion. The crucial ingredient to realize Poincar\'{e}'s dream of
complete reduction of dynamics to fibers (i.e. to $\mathfrak{g}$ or $%
\mathfrak{g}^{\ast }$) is the dimensions of $G$ and $N$. For second order
Lagrangians, replacing $N$ with $TM$ changes this dimension count$.$ Since,
for chiral oscillator 
\begin{equation*}
\dim \mathfrak{se}(2)=3<4=\dim TM=2\dim M,
\end{equation*}%
we still need an additional symmetry for a complete reduction. Surprisingly,
this additional symmetry comes from a central extension of symmetry
algebra.\ \ 

\section{Hamiltonian structures}

Dirac-Bergman theory of constraints identifies final constraint submanifold
of a degenerate Lagrangian theory on which Dirac-Hamilton structure is
nondegenerate \cite{dirac}-\cite{gotay80}. This is combined with
Ostrogradskii's theory of Hamiltonian formulations for degenerate higher
derivative Lagrangians \cite{batlle88}-\cite{gracia91},\cite{hg22}.

The underlying geometric structures of chiral oscillator Lagrangian along
with general descriptions of Dirac's and Ostrogradskii's methods for second
order degenerate Lagrangians are studied in \cite{luk}-\cite{cruz} and \cite%
{hg22}.

Starting with the variation of $L_{Chiral}\left[ x,y\right] $ we have the
Euler-Lagrange equations 
\begin{equation}
\lambda \dddot{y}-m\ddot{x}=0\text{, \ \ \ \ \ }-\lambda \dddot{x}-m\ddot{y}%
=0  \label{ele}
\end{equation}%
and the Ostrogradskii momenta $\mathbf{p}^{0}=(p_{x}^{0},p_{y}^{0})$ and $%
\mathbf{p}^{1}=(p_{x}^{1},p_{y}^{1})$ with components 
\begin{equation}
p_{x}^{0}=-\lambda \ddot{y}+m\dot{x}\text{, \ }p_{y}^{0}=\lambda \ddot{x}+m%
\dot{y}\text{, \ }p_{x}^{1}=\frac{\lambda }{2}\dot{y}\text{, \ }p_{y}^{1}=-%
\frac{\lambda }{2}\dot{x}  \label{mom12}
\end{equation}%
the last two of which define the second class constraints%
\begin{equation}
\phi _{x}=\ p_{x}^{1}-\lambda \dot{y}/2,\ \phi _{y}=p_{y}^{1}+\lambda \dot{x}%
/2,\ \ \ \ \left\{ \phi _{x},\phi _{y}\right\} _{C}=-\lambda  \label{const}
\end{equation}%
where we used the canonical structure on $T^{\ast }TM$ with coordinates\ $%
\mathbf{z:=}(\mathbf{x},\mathbf{\dot{x}},\mathbf{p}^{0},\mathbf{p}^{1})\in
T^{\ast }TM$ and the Poisson brackets%
\begin{equation}
\left\{ x,p_{x}^{0}\right\} _{C}=\left\{ y,p_{y}^{0}\right\} _{C}=\left\{ 
\dot{x},p_{x}^{1}\right\} _{C}=\left\{ \dot{y},p_{x}^{1}\right\} _{C}=1
\label{cbr}
\end{equation}%
The Dirac bracket can be found to have the matrix \cite{hg22}%
\begin{equation}
P_{D}(\mathbf{z})=\left( 
\begin{array}{cccc}
0 & 0 & I_{2} & 0 \\ 
0 & -\lambda J_{2}/2 & 0 & I_{2}/2 \\ 
-I_{2} & 0 & 0 & 0 \\ 
0 & -I_{2}/2 & 0 & \lambda J_{2}/4%
\end{array}%
\right)  \label{dmat}
\end{equation}%
where the $2\times 2$ identity and symplectic submatrices are 
\begin{equation*}
\text{\ \ }I_{2}=\left( 
\begin{array}{cc}
1 & 0 \\ 
0 & 1%
\end{array}%
\right) ,\text{ \ \ \ }J_{2}=\left( 
\begin{array}{cc}
0 & 1 \\ 
-1 & 0%
\end{array}%
\right) .
\end{equation*}%
In particular, the Dirac brackets of coordinate functions are%
\begin{eqnarray}
\left\{ x,p_{x}^{0}\right\} _{D} &=&\left\{ y,p_{y}^{0}\right\} _{D}=1,\text{
\ }\left\{ \dot{x},\dot{y}\right\} _{D}=\frac{1}{\lambda }  \label{dbr1} \\
\left\{ \dot{x},p_{x}^{1}\right\} _{D} &=&\left\{ \dot{y},p_{x}^{1}\right\}
_{D}=\frac{1}{2},\text{ \ \ }\left\{ p_{x}^{1},p_{y}^{1}\right\} _{D}=\frac{%
\lambda }{4}.  \label{dbr2}
\end{eqnarray}%
The Hamilton's equations can then be written as%
\begin{equation}
\mathbf{\dot{z}}=P_{D}(\mathbf{z})\nabla _{z}H^{C}(\mathbf{z})  \label{direq}
\end{equation}%
where $H^{C}$ is the canonical Hamiltonian function%
\begin{equation}
H^{C}(\mathbf{z})=p_{x}^{0}\dot{x}+p_{y}^{0}\dot{y}-\frac{m}{2}(\dot{x}^{2}+%
\dot{y}^{2})=\mathbf{\dot{x}}\cdot \mathbf{p}^{0}-\frac{m}{2}\mid \mathbf{%
\dot{x}}\mid ^{2}.  \label{cham}
\end{equation}%
Eq.(\ref{direq}) consists of definition and conservation law for linear
momenta $\mathbf{p}^{0}$ and identities. Equivalently, the total
Hamiltonian, that is the Hamiltonian function obtained from the
Dirac-Bergman analysis%
\begin{eqnarray}
H^{D}(\mathbf{z}) &=&\frac{1}{2}(p_{x}^{0}\dot{x}+p_{y}^{0}\dot{y})-\frac{m}{%
\lambda }(p_{x}^{1}\dot{y}-p_{y}^{1}\dot{x})-\frac{1}{\lambda }%
(p_{x}^{0}p_{y}^{1}-p_{y}^{0}p_{x}^{1}) \\
&=&\frac{1}{2}\mathbf{\dot{x}}\cdot \mathbf{p}^{0}+\frac{m}{\lambda }\mathbf{%
\dot{x}}\times \mathbf{p}^{1}-\frac{1}{\lambda }\mathbf{p}^{0}\times \mathbf{%
p}^{1}  \label{dh}
\end{eqnarray}%
can also be used, together with canonical Poisson bracket, to write
Hamilton's equations%
\begin{equation*}
\mathbf{\dot{z}}=P_{C}(\mathbf{z})\nabla _{z}H^{D}(\mathbf{z})
\end{equation*}%
which manifestly read%
\begin{eqnarray}
\dot{x} &=&-\frac{2}{\lambda }p_{y}^{1}\text{, \ \ }\dot{y}=\frac{2}{\lambda 
}p_{x}^{1}\text{,\ \ \ }\ddot{x}=-\frac{m}{\lambda }\dot{y}+\frac{1}{\lambda 
}p_{y}^{0}\text{, \ \ }\ddot{y}=\frac{m}{\lambda }\dot{x}-\frac{1}{\lambda }%
p_{x}^{0}  \label{he1} \\
\dot{p}_{x}^{0} &=&0\text{,\ \ \ }\dot{p}_{y}^{0}=0\text{,\ \ \ \ }\dot{p}%
_{x}^{1}=-\frac{m}{\lambda }p_{y}^{1}-\frac{1}{2}p_{x}^{0}\text{, \ \ }\dot{p%
}_{y}^{1}=\frac{m}{\lambda }p_{x}^{1}-\frac{1}{\lambda }p_{y}^{0}\text{.}
\label{he2}
\end{eqnarray}%
Apart from definitions of Ostrogradskii momenta in the first line, these
equations again reduce Euler-Lagrange equations to conservation laws for
linear momenta $p_{x}^{0},p_{y}^{0}$.

Yet another form of Hamilton's equations can be obtained by bringing the
Dirac bracket into canonical form on six dimensional final constrained
submanifold. This can be achieved by introducing the Darboux coordinates
(see e.g. \cite{am},\cite{mr},\cite{marle14}) for Dirac%
\begin{equation}
\mathbf{w}(\mathbf{z})=\left( x,y,q,p_{x}^{0},p_{y}^{0},p)\oplus (\phi _{x},%
\text{ }\phi _{y}\right) .  \label{wz}
\end{equation}%
New coordinates factor out the constraints and the submanifold on which the
dynamics takes place. We will denote this final constrained submanifold by $%
T^{\ast }D_{f}$ as it is endowed with canonically conjugate coordinates $(%
\mathbf{x},q;\mathbf{p}^{0},p)$ where%
\begin{eqnarray}
q &=&\dot{x}+\frac{2}{\lambda }p_{y}^{1}+\frac{2}{\sqrt{\lambda }}p_{x}^{1}=%
\sqrt{\lambda }\dot{y},\text{ }  \label{darq} \\
p &=&\dot{y}-\frac{2}{\lambda }p_{x}^{1}+\frac{2}{\sqrt{\lambda }}p_{y}^{1}=-%
\sqrt{\lambda }\dot{x}.  \label{darp}
\end{eqnarray}%
The transformation $\mathbf{w}(\mathbf{z})$ leads the Ostrogradskii phase
space to admit symplectic orthogonal decomposition%
\begin{equation*}
(T^{\ast }TM,\omega _{C})\rightsquigarrow (T^{\ast }D_{f},\omega _{f})\oplus
(T^{\ast }%
\mathbb{R}
,d\phi _{x}\wedge d\phi _{y})
\end{equation*}%
where the canonical symplectic two-form on the final constrained submanifold 
$T^{\ast }D_{f}$ is 
\begin{equation}
\omega _{f}=-dp_{x}^{0}\wedge dx-dp_{y}^{0}\wedge dy-dp\wedge dq\text{.}
\label{omf}
\end{equation}%
Equivalently, the Poisson bracket relations on $T^{\ast }TM$ reduces to 
\begin{equation}
\left\{ x,p_{x}^{0}\right\} _{f}=\left\{ y,p_{y}^{0}\right\} _{f}=\left\{
q,p\right\} _{f}=1.  \label{fbra}
\end{equation}%
The pull-back of $\omega _{f}$ by $\mathbf{w}(\mathbf{z})$ gives the
Lagrangian symplectic two-form $\omega _{L}$ of $L_{Chiral}$. In other
words, the Legendre transformation $\mathbb{F}L:T^{3}M\longrightarrow
T^{\ast }TM$ for a general second order Lagrangian \cite{batlle88}-\cite%
{gracia91} becomes a diffeomorphism 
\begin{equation*}
\mathbb{F}L_{D_{f}}=\mathbb{F}L\mid _{\phi _{x}=\phi
_{y}=0}:T^{2}M\longrightarrow T^{\ast }D_{f}
\end{equation*}%
when restricted to final constrained submanifold of the chiral oscillator
Lagrangian. Moreover, using pull-back $\mathbb{F}L_{D_{f}}^{\ast }$ by
Legendre map, we have 
\begin{equation*}
\mathbb{F}L_{D_{f}}^{\ast }\omega _{f}=\omega _{L}\text{, \ \ \ \ \ }%
i_{X_{L}}\omega _{L}=-dE_{L}=-d\mathbb{F}L_{D_{f}}^{\ast }H_{f}\text{ ,\ \ \
\ \ }
\end{equation*}%
where $E_{L}$ is the energy function%
\begin{equation*}
E_{L}=-\lambda (\dot{x}\ddot{y}-\dot{y}\ddot{x})+\frac{m}{2}(\dot{x}^{2}+%
\dot{y}^{2}),
\end{equation*}%
$\omega _{L}$ is the Lagrangian symplectic two-form%
\begin{equation*}
\omega _{L}\left[ x,y\right] =-d\theta _{L}\left[ x,y\right] =\frac{\lambda 
}{2}d\dot{x}\wedge d\dot{y}+\left( \lambda d\ddot{y}-md\dot{x}\right) \wedge
dx-\left( \lambda d\ddot{x}+md\dot{y}\right) \wedge dy
\end{equation*}%
and $H_{f}$ is the canonical Hamiltonian 
\begin{equation}
H_{f}=\frac{1}{\sqrt{\lambda }}(qp_{y}^{0}-pp_{x}^{0})-\frac{m}{2\lambda }%
(q^{2}+p^{2})  \label{fham}
\end{equation}%
expressed in Darboux coordinates of $T^{\ast }D_{f}.$ The composition of
Hamilton's equations is identical with previous structures.

Final constrained submanifold $T^{\ast }D_{f}$ endowed with its canonical
coordinates can be used to give a vector bundle structure to $T^{2}M$ via
inverse Legendre transform 
\begin{equation*}
\mathbb{F}L_{D_{f}}^{-1}=T^{\ast }D_{f}\longrightarrow T^{2}M=:TD_{f}
\end{equation*}%
\begin{equation}
\dot{x}=-\frac{p}{\sqrt{\lambda }},\text{ \ \ }\dot{q}=-\frac{p_{x}^{0}}{%
\sqrt{\lambda }}-\frac{m}{\lambda }p,\text{ \ \ \ }\dot{y}=\frac{q}{\sqrt{%
\lambda }}  \label{legtr}
\end{equation}%
which realizes the chiral oscillator Lagrangian as a first order
nondegenerate constrained Lagrangian 
\begin{equation}
L_{f}\left[ x,y,q\right] =\frac{m}{2}\dot{x}^{2}-\sqrt{\lambda }\dot{x}\dot{q%
}+p_{y}^{0}(\dot{y}-\frac{q}{\sqrt{\lambda }})+\frac{m}{2\lambda }q^{2}
\label{flag}
\end{equation}%
with Lagrange multiplier $p_{y}^{0},$ over three dimensional configuration
manifold $D_{f}$. Attempt to make $L_{f}$ into an unconstraint Lagrangian,
that is, to solve Lagrange multiplier $p_{y}^{0}$ turns $L_{f}$ back into
the second order chiral Lagrangian $L_{Chiral}$. See Horv\'{a}thy and
Plyushchay \cite{hor02} for a similar first order constrained Lagrangian
over three dimensional configuration space equivalent to $L_{Chiral}$.

\section{Symmetries and oscillator algebra}

General treatment, within the context of reductions, of $SE(2)$ along with
its\ central extension as well as discussions on\ oscillator algebra can be
found in \cite{bigstages07} and \cite{vkm10}.

The group $SE(2)$ of Euclidean motions of plane $M$ consists of rotation $%
SO(2)$ and translations $%
\mathbb{R}
^{2}$. It is a semidirect product $SO(2)\ltimes 
\mathbb{R}
^{2}$ with rotations acting on translations. Group of translations is
Abelian and is a normal subgroup that corresponds to an ideal in Lie algebra 
$\mathfrak{se}(2)$. For elements parametrized as 
\begin{equation*}
\text{\ }R_{\theta }=\left( 
\begin{array}{cc}
\cos \theta & -\sin \theta \\ 
\sin \theta & \cos \theta%
\end{array}%
\right) \in SO(2)\text{, \ \ }\mathbf{a}=\left( 
\begin{array}{c}
a \\ 
b%
\end{array}%
\right) \in 
\mathbb{R}
^{2}\text{, \ }
\end{equation*}%
the action of $SE(2)$ on $M$ is given by%
\begin{equation*}
\Phi _{(R_{\theta },\mathbf{a)}}(\mathbf{x)}\mathbf{=}R_{\theta }\mathbf{x}+%
\mathbf{a}\text{, \ \ \ \ \ \ }\mathbf{x}=\left( 
\begin{array}{c}
x \\ 
y%
\end{array}%
\right) \in M.
\end{equation*}%
The generators of this action in coordinates of $M$ are 
\begin{equation}
R(x,y)=x\frac{\partial }{\partial y}-y\frac{\partial }{\partial x},\text{ \
\ }X(x,y)=\frac{\partial }{\partial x},\text{ \ \ }Y(x,y)=\frac{\partial }{%
\partial y}\text{\ }\in T_{(x,y)}M  \label{seg}
\end{equation}%
and they satisfy the Lie bracket relations%
\begin{equation}
\left[ R,X\right] =Y,\text{ \ \ }\left[ R,Y\right] =-X,\text{ \ \ }\left[ X,Y%
\right] =0  \label{lb}
\end{equation}%
of the Lie algebra $\mathfrak{se}(2).$ This Lie algebra acts on $TM$ by
tangent lifts%
\begin{equation}
F_{R}^{TM}=R(x,y)+\dot{x}\frac{\partial }{\partial \dot{y}}-\dot{y}\frac{%
\partial }{\partial \dot{x}},\text{ \ \ }F_{X}^{TM}=X(x,y),\text{ \ \ }%
F_{Y}^{TM}=Y(x,y)  \label{tanlift}
\end{equation}%
and on $T^{\ast }TM$ by cotangent lifts of tangent lifts%
\begin{equation}
F_{R}^{T^{\ast }TM}=F_{R}^{TM}+p_{x}^{0}\frac{\partial }{\partial p_{y}^{0}}%
-p_{y}^{0}\frac{\partial }{\partial p_{x}^{0}}+p_{x}^{1}\frac{\partial }{%
\partial p_{y}^{1}}-p_{y}^{1}\frac{\partial }{\partial p_{x}^{1}},
\label{cotlift}
\end{equation}%
with $F_{X}^{T^{\ast }TM}=X(x,y)$ and\ $F_{Y}^{T^{\ast }TM}=Y(x,y)$ still
being the same. These lifts are Hamiltonian vector fields with the canonical
symplectic structure of $T^{\ast }TM$ for the Hamiltonian functions%
\begin{equation}
\mu =\mathbf{x}\times \mathbf{p}^{0}+\mathbf{\dot{x}}\times \mathbf{p}^{1}%
\text{, \ \ }p_{x}^{0}\text{, \ \ }p_{y}^{0}  \label{mup0}
\end{equation}%
which are angular and linear momenta and constitute coordinates of the dual
space $\mathfrak{se}(2)^{\ast }$ to symmetry algebra. These are the
components of momentum map $\mathbb{J}^{T^{\ast }TM}:T^{\ast }TM\rightarrow 
\mathfrak{se}(2)^{\ast }$ for the action generated by cotangent lifts. For
the action on $T^{\ast }M$ endowed with the canonical symplectic form $d%
\mathbf{p}^{0}\wedge d\mathbf{x}$, the momentum map is%
\begin{equation}
\mathbb{J}^{T^{\ast }M}:T^{\ast }M\rightarrow \mathfrak{se}(2)^{\ast }\text{
\ \ \ \ \ \ }\mathbb{J}^{T^{\ast }M}(\mathbf{x},\mathbf{p}^{0})=(\mathbf{x}%
\times \mathbf{p}^{0},\mathbf{p}^{0}).  \label{p0}
\end{equation}%
Yet another momentum map is associated with the action of $SE(2)$ on $M$
itself as a submanifold of the symplectic space $(%
\mathbb{R}
^{2},dx\wedge dy)$. This is also the symplectic structure on $SE(2)$
defining a two-cocyle involving its central extension. In this case, we have%
\begin{equation}
\mathbb{J}^{M}:M\subseteq 
\mathbb{R}
^{2}\rightarrow \mathfrak{se}(2)^{\ast }\text{ \ \ \ \ \ \ }\mathbb{J}^{M}(%
\mathbf{x})=(\frac{1}{2}\mid \mathbf{x}\mid ^{2},y,-x).  \label{momplane}
\end{equation}

The coadjoint action of $SE(2)$ on dual of its Lie algebra is by translating
the angular and rotating the linear momenta. Evaluation of canonical
brackets of momenta (\ref{mup0}) on $T^{\ast }TM$ or, momenta (\ref{p0}) on $%
T^{\ast }M$, gives $\mathfrak{se}(2)^{\ast }$ Lie-Poisson brackets 
\begin{equation*}
\left( 
\begin{array}{ccc}
0 & -p_{y}^{0} & p_{x}^{0} \\ 
p_{y}^{0} & 0 & 0 \\ 
-p_{x}^{0} & 0 & 0%
\end{array}%
\right) .
\end{equation*}%
Being three dimensional, this Poisson structure is degenerate and possesses
the Casimir function%
\begin{equation}
l^{2}:=(p_{x}^{0})^{2}+(p_{y}^{0})^{2}=\mid \mathbf{p}^{0}\mid ^{2}
\label{cas}
\end{equation}%
which is the length-squared of linear momenta.

Hamiltonian reductions resemble this construction of Lie-Poisson brackets
via evaluation of canonical brackets of components of momentum map.
Hamiltonian function governing dynamics of particular system under
consideration may require more general invariants than components of
momentum maps for cotangent lifted action which are only linear functions of
fiber coordinates. Indeed, they are invariants on $T^{\ast }TM$ for which
momentum map is a Poisson morphism.

\subsection{Central extension and oscillator algebra}

The commuting generators $X$ and $Y$ of translations are Hamiltonian vector
fields for the symplectic space $(M\subseteq 
\mathbb{R}
^{2},dx\wedge dy)$ and with the Hamiltonian functions $J_{X}^{M}=-y$ and $%
J_{Y}^{M}=x$, respectively. These are the components of the momentum map in
Eq.(\ref{momplane}) for the action of $SE(2)$ on $M$ and satisfy $\left\{
J_{X}^{M},J_{Y}^{M}\right\} _{%
\mathbb{R}
^{2}}=1$. On the other hand, since $\left[ X,Y\right] =0$ we have $J_{\left[
X,Y\right] }^{M}=0$. Define the bilinear function $\Theta _{%
\mathbb{R}
^{2}}:\mathfrak{se(}2)\times \mathfrak{se(}2)\rightarrow 
\mathbb{R}
$ by its values%
\begin{equation}
\Theta _{%
\mathbb{R}
^{2}}(X,Y)=\left\{ J_{X}^{M},J_{Y}^{M}\right\} _{%
\mathbb{R}
^{2}}-J_{\left[ X,Y\right] }^{M}=1\text{, \ \ }\Theta _{%
\mathbb{R}
^{2}}(Y,R)=\Theta _{%
\mathbb{R}
^{2}}(R,X)=0.  \label{cocycle}
\end{equation}%
$\Theta _{%
\mathbb{R}
^{2}}$ satisfies the two-cocycle identity%
\begin{equation*}
\Theta _{%
\mathbb{R}
^{2}}(\left[ X,Y\right] ,R)+\Theta _{%
\mathbb{R}
^{2}}(\left[ Y,R\right] ,X)+\Theta _{%
\mathbb{R}
^{2}}(\left[ R,X\right] ,Y)=0
\end{equation*}%
automatically. Thus, breaking of homomorhism between Lie bracket algebra of
Hamiltonian vector fields and Poisson bracket algebra of \ Hamiltonian
functions in the first equation of (\ref{cocycle}) is connected to the
existence of Lie algebra two-cocycle. The situation can also be related to
non-equivariance of momentum map. See \cite{am},\cite{mr},\cite{marle14},%
\cite{cmr04} for details. \cite{vkm10} contains explicit computation of $%
\mathfrak{se(}2)$ two-cocycle for the above action on $M.$

In the present context of second order Lagrangians, our concern is the
action of $\mathfrak{se(}2)$ on $T^{\ast }TM$. However, cotangent lifts of
tangent lifts of generators $X$ and $Y$ remains to be themselves. That means
the bilinear function defining two-cocycle is still a constant. We will see
that this constant can be extended to Casimir function (\ref{cas}) of $%
\mathfrak{se(}2)$ when more appropriate lifts of action involve.

In order to preserve the Lie algebra homomorphism between Hamiltonian vector
fields and functions as well as to reserve equivariance of momentum map, one
considers the central extension of symmetry algebra $\mathfrak{se(}2)$ by
the two-cocycle $\Theta _{%
\mathbb{R}
^{2}}$. \ The result is the four dimensional algebra defined by the Lie
bracket relations%
\begin{equation}
\left[ R,X\right] =Y,\text{ \ \ }\left[ R,Y\right] =-X,\text{ \ \ }\left[ X,Y%
\right] =Z_{\Theta _{%
\mathbb{R}
^{2}}(X,Y)}  \label{osc}
\end{equation}%
where $Z$ is, in general, the Hamiltonian vector field defined by the
function $\Theta _{%
\mathbb{R}
^{2}}(X,Y)$. Streater \cite{streater67} coined the name harmonic oscillator
algebra for reason that in proving Jacobi identities, he used analogy with
the generators%
\begin{equation*}
R_{ho}=-\frac{\partial ^{2}}{\partial x^{2}}+x^{2}\text{, \ \ }X_{ho}=x\text{%
, \ \ }Y_{ho}=\frac{\partial }{\partial x}\text{, \ \ }Z_{ho}=1
\end{equation*}%
associated to harmonic oscillator problem.

The Lie-Poisson bracket on $\mathfrak{osc}^{\ast }$ has the matrix 
\begin{equation*}
\left( 
\begin{array}{cccc}
0 & -p_{y} & p_{x} & 0 \\ 
p_{y} & 0 & -\Theta _{%
\mathbb{R}
^{2}}(X,Y) & 0 \\ 
-p_{x} & \Theta _{%
\mathbb{R}
^{2}}(X,Y) & 0 & 0 \\ 
0 & 0 & 0 & 0%
\end{array}%
\right)
\end{equation*}%
in the coordinates $(\mu ,p_{x},p_{y},\Theta _{%
\mathbb{R}
^{2}}(X,Y))$. More generally, for actions on various symplectic spaces with
momentum map $\mathbb{J}=(J_{R},J_{X},J_{Y})$ we have the Lie-Poisson
bracket relations%
\begin{equation}
\left\{ J_{R},J_{X}\right\} _{\mathfrak{osc}^{\ast }}=J_{Y},\text{ \ \ }%
\left\{ J_{R},J_{Y}\right\} _{\mathfrak{osc}^{\ast }}=-J_{X},\text{ \ \ }%
\left\{ J_{X},J_{Y}\right\} _{\mathfrak{osc}^{\ast }}=\Theta _{%
\mathbb{R}
^{2}}(X,Y).  \label{pbo}
\end{equation}%
As noted in \cite{vkm10} the projection $\mathfrak{osc}^{\ast }\rightarrow 
\mathfrak{se(}2)^{\ast }$ is a Poisson map.

\section{Cotangent bundle reductions}

Reduction of Hamiltonian systems that are equivalent to the Euler-Lagrange
equations for the chiral oscillator Lagrangian will be obtained by
evaluation of corresponding Poisson brackets, namely, canonical bracket (\ref%
{cbr}) on $T^{\ast }TM$, Dirac bracket (\ref{dbr1}),(\ref{dbr2}) on $T^{\ast
}TM$ and canonical-Dirac bracket (\ref{fbra}) on $T^{\ast }D_{f}$, for
functions on $\mathbb{J}^{-1}(\nu )$ with the following properties: a) They
are invariant under $SE(2).$ b) In each case, they make up the corresponding
Hamiltonian function for chiral oscillator dynamics. c) They parametrize
(form coordinates on) $\mathbb{J}^{-1}(\nu ).$ First requirement implies
that these functions must be independent of $x$ and $y$ for translation
invariance and, be quadratic for rotational invariance. For reduction of
Dirac bracket, in particular, the one on $T^{\ast }D_{f},$ this first
requirement alone gives four functions%
\begin{equation*}
\mid \mathbf{\dot{x}}\mid ^{2},\text{ \ }\mid \mathbf{p}^{0}\mid ^{2},\text{
\ }\mathbf{\dot{x}}\cdot \mathbf{p}^{0},\text{ \ }\mathbf{\dot{x}}\times 
\mathbf{p}^{0}
\end{equation*}%
which is the only choice of invariant functions, thereby showing the
uniqueness of oscillator algebra structure on reduced space. For regularity
conditions, we will assume $\mathbf{p}^{0}\neq 0$ and $\mathbf{\dot{x}}%
\times \mathbf{p}^{1}\neq 0$.

\subsection{Canonical bracket with Dirac Hamiltonian}

Define $SE(2)$ invariant functions%
\begin{eqnarray}
J_{R} &=&-p_{x}^{1}\dot{y}+p_{y}^{1}\dot{x}=\mathbf{\dot{x}}\times \mathbf{p}%
^{1}\text{ \ \ \ \ \ \ \ \ \ }  \label{jr} \\
J_{X} &=&\frac{1}{2}(p_{x}^{0}\dot{x}+p_{y}^{0}\dot{y})-\frac{1}{\lambda }%
(p_{x}^{0}p_{y}^{1}-p_{y}^{0}p_{x}^{1})=\frac{1}{2}\mathbf{\dot{x}}\cdot 
\mathbf{p}^{0}-\frac{1}{\lambda }\mathbf{p}^{0}\times \mathbf{p}^{1}
\label{jx} \\
J_{Y} &=&\frac{1}{2}(p_{x}^{0}\dot{y}-p_{y}^{0}\dot{x})+\frac{1}{\lambda }%
(p_{x}^{0}p_{x}^{1}+p_{y}^{0}p_{y}^{1})=-\frac{1}{2}\mathbf{\dot{x}}\times 
\mathbf{p}^{0}+\frac{1}{\lambda }\mathbf{p}^{0}\cdot \mathbf{p}^{1}
\label{jy}
\end{eqnarray}%
on $T^{\ast }TM.$ The Dirac's total Hamiltonian function in Eq.(\ref{dh})
becomes%
\begin{equation}
\text{\ }H_{red}^{D}=\frac{m}{\lambda }J_{R}+J_{X}.  \label{rdh}
\end{equation}%
With the canonical bracket (\ref{cbr}) on $T^{\ast }TM$ we find that they
satisfy the Poisson bracket algebra (\ref{pbo}) of $\mathfrak{osc}^{\ast }$%
\begin{equation}
\left\{ J_{R},J_{X}\right\} _{C}=J_{Y},\text{ \ \ }\left\{
J_{R},J_{Y}\right\} _{C}=-J_{X},\text{ \ \ }\left\{ J_{X},J_{Y}\right\} _{C}=%
\frac{1}{\lambda }l^{2}  \label{osclp}
\end{equation}%
with Casimir function $l^{2}$ commuting with all others. That is, in
coordinates $(J_{R},J_{X},J_{Y},l^{2}/\lambda )$ of $\mathfrak{osc}^{\ast }$
(literally, in coordinates on the inverse image of a momentum map $T^{\ast
}TM\longrightarrow $ $\mathfrak{osc}^{\ast }$ for action of oscillator
group) we have the Poisson matrix 
\begin{equation*}
\left( 
\begin{array}{cccc}
0 & J_{Y} & -J_{X} & 0 \\ 
-J_{Y} & 0 & l^{2}/\lambda & 0 \\ 
J_{X} & l^{2}/\lambda & 0 & 0 \\ 
0 & 0 & 0 & 0%
\end{array}%
\right) .
\end{equation*}%
Using reduced Hamiltonian function (\ref{rdh}) we obtain the reduced
equations 
\begin{equation}
\dot{J}_{R}=J_{Y},\text{ \ \ }\dot{J}_{X}=-\frac{m}{\lambda }J_{Y},\text{ \
\ }\dot{J}_{Y}=\frac{m}{\lambda }J_{X}-\frac{1}{\lambda }l^{2},\text{ \ \ }%
\dot{l}=0  \label{lpcan}
\end{equation}%
which can be shown to be equivalent to Hamilton's equations (\ref{he1}),(\ref%
{he2}) via definitions (\ref{jr})-(\ref{jy}) of invariant functions provided
the regularity conditions are satisfied. By adding and subtracting the term $%
m(p_{x}^{1}\dot{x}+p_{y}^{1}\dot{y})/\lambda $, first equation can be
brought to a form where Hamilton's equations are coefficients of coordinates 
$(\dot{x},\dot{y},p_{x}^{1},p_{y}^{1})$ on phase space. Similarly, second
and third equations displays Hamilton's equations as coefficients of momenta 
$(p_{x}^{0},p_{y}^{0})$. The last one is the conservation laws for the
momenta $(p_{x}^{0},p_{y}^{0})$.

The last equation in (\ref{lpcan}) means that the reduced dynamics is
defined on Casimir surfaces of $\mathfrak{se}(2)^{\ast }$. As for any
Poisson structure in three dimensions, the remaining three equations admit
the Casimir function 
\begin{equation*}
C=J_{X}^{2}-2ml^{2}J_{X}+J_{Y}^{2}
\end{equation*}%
which, together with $H_{red}^{D}$ form characteristics of reduced
equations. In other words, solutions of the reduced equations (\ref{osclp})
are intersections of level sets of $H_{red}^{D}$ and $C$ on level sets of $%
l^{2}$.

\subsection{Dirac bracket with canonical Hamiltonian}

As Dirac-Bergman procedure eliminates the momenta $\mathbf{p}^{1}$ through
constraints, we have, in this case, the $SE(2)$ invariant functions 
\begin{eqnarray}
\text{\ }J_{R}^{D} &=&-\frac{\lambda }{2}(\dot{x}^{2}+\dot{y}^{2})=-\text{\ }%
\frac{\lambda }{2}\mid \mathbf{\dot{x}}\mid ^{2}\text{,}  \label{jrd} \\
\text{\ }J_{X}^{D} &=&\dot{x}p_{x}^{0}+\dot{y}p_{y}^{0}\mathfrak{=}\mathbf{%
\dot{x}}\cdot \mathbf{p}^{0}\text{, \ \ \ }  \label{jxd} \\
J_{Y}^{D} &=&-(\dot{x}p_{y}^{0}-\dot{y}p_{x}^{0})=-\mathbf{\dot{x}}\times 
\mathbf{p}^{0}  \label{jyd}
\end{eqnarray}%
on $T^{\ast }TM$ which, together with $l^{2},$ form unique coordinates to
parametrize reduced Hamiltonian space $\mathbb{J}^{-1}(\nu )$. With this
parametrization, the Dirac bracket (\ref{dbr1}),(\ref{dbr2}) on $T^{\ast }TM$
reduces to Lie-Poisson bracket (\ref{pbo}) of $\mathfrak{osc}^{\ast }$ as in
Eq.(\ref{osclp}) together with the usual vanishing commutators of $l^{2}$.
The functions $J_{R}^{D},J_{X}^{D}$ and $J_{Y}^{D}$ are Hamiltonian
function, with respect to Dirac bracket, for the following lifts%
\begin{eqnarray*}
F_{R}^{T^{\ast }TM} &=&\dot{x}\frac{\partial }{\partial \dot{y}}-\dot{y}%
\frac{\partial }{\partial \dot{x}}, \\
F_{X}^{T^{\ast }TM} &=&\dot{x}\frac{\partial }{\partial x}+\dot{y}\frac{%
\partial }{\partial y}+\frac{1}{\lambda }(p_{y}^{0}\frac{\partial }{\partial 
\dot{x}}-p_{x}^{0}\frac{\partial }{\partial \dot{y}}), \\
F_{Y}^{T^{\ast }TM} &=&\dot{y}\frac{\partial }{\partial x}-\dot{x}\frac{%
\partial }{\partial y}+\frac{1}{\lambda }(p_{x}^{0}\frac{\partial }{\partial 
\dot{x}}+p_{y}^{0}\frac{\partial }{\partial \dot{y}})
\end{eqnarray*}%
to $T^{\ast }TM$ of $\mathfrak{se}(2)$ Lie algebra generators $R,X$ and $Y$
in Eq.(\ref{seg}), respectively. The first one is obviously part of rotation
acting on velocities. Together with the Hamiltonian vector field%
\begin{equation*}
F_{l^{2}/\lambda }^{T^{\ast }TM}=\frac{2}{\lambda }(p_{x}^{0}\frac{\partial 
}{\partial x}+p_{y}^{0}\frac{\partial }{\partial y})
\end{equation*}%
for the Casimir function (\ref{cas}) with respect to Dirac bracket again,
they satisfy the Lie bracket relations 
\begin{equation*}
\left[ F_{R}^{T^{\ast }TM},F_{X}^{T^{\ast }TM}\right] =-F_{Y}^{T^{\ast }TM},%
\text{ \ }\left[ F_{R}^{T^{\ast }TM},F_{Y}^{T^{\ast }TM}\right]
=F_{X}^{T^{\ast }TM},\text{ \ }\left[ F_{X}^{T^{\ast }TM},F_{Y}^{T^{\ast }TM}%
\right] =-F_{l^{2}/\lambda }^{T^{\ast }TM}
\end{equation*}%
of oscillator algebra in Eq.(\ref{osc}) with $F_{l^{2}/\lambda }^{T^{\ast
}TM}$ commuting with all others. Thus, the appropriate two-cocycle for
chiral oscillator is defined by the Casimir function of $\mathfrak{se}%
(2)^{\ast }$ as $\Theta _{T^{\ast }TM}(\mathbf{z})=l^{2}/\lambda $.

In terms of the invariants (\ref{jrd})-(\ref{jyd}), the reduced canonical
Hamiltonian function becomes%
\begin{equation}
H_{red}^{C}=\dot{x}p_{x}^{0}+\dot{y}p_{y}^{0}-\frac{m}{2}(\dot{x}^{2}+\dot{y}%
^{2})=\frac{m}{\lambda }J_{R}^{D}+J_{X}^{D}  \label{hdred}
\end{equation}%
and the reduced Hamilton's equations are as in Eq.(\ref{lpcan}) with new and
simpler invariants. The first equation in (\ref{lpcan}) is the conservation
of angular momentum $\mu =\mathbf{x}\times \mathbf{p}^{0}+J_{R}^{D}$ and the
second equation is identically satisfied when the last equation, the
conservation of linear momenta, along with its definition used. More
precisely, we compute%
\begin{eqnarray*}
\lambda \dot{J}_{X}^{D}+mJ_{Y}^{D} &=&\lambda \mathbf{\ddot{x}}\cdot \mathbf{%
p}^{0}-m\mathbf{\dot{x}}\times \mathbf{p}^{0} \\
&=&\lambda (\ddot{x}p_{x}^{0}+\ddot{y}p_{y}^{0})-m(\dot{x}p_{y}^{0}-\dot{y}%
p_{x}^{0}) \\
&=&\mathbf{p}^{0}\times \mathbf{p}^{0}=0.
\end{eqnarray*}%
It follows from the defining relations (\ref{jrd})-(\ref{jyd}) that the
coordinates on the reduced space satisfy%
\begin{equation}
(J_{X}^{D})^{2}+(J_{Y}^{D})^{2}+\frac{2l^{2}}{\lambda }J_{R}^{D}=0
\label{paraboloid}
\end{equation}%
which, for each regular value of the Casimir $l^{2}$, a circular parabaloid 
\cite{vkm10}. These two dimensional surfaces are conclusion of
Marsden-Weinstein reduction theorem \cite{mw}-\cite{mr}: they are symplectic
leaves of the reduced Poisson structure with the symplectic two-form 
\begin{equation}
\omega _{\mathcal{O}}=\frac{\lambda }{l^{2}}dJ_{Y}^{D}\wedge dJ_{X}^{D}
\label{kk}
\end{equation}%
where $\mathcal{O}$ refers to coadjoint orbit symplectic structure.

Reduction in Dirac bracket and canonical Hamiltonian function in Darboux
coordinates turns out to be identical with the above reduction. Canonical
Dirac bracket on final constrained submanifold $T^{\ast }D_{f}$ of $T^{\ast
}TM$ can be obtained from transformation in Eqs.(\ref{wz}-\ref{darp}) to
Darboux coordinates. If $(\partial \mathbf{w}/\partial \mathbf{z})$ denotes
the Jacobian matrix of this transformation, then the matrix of Dirac's
bracket in new coordinates will be%
\begin{equation*}
P_{D}(\mathbf{w})=\left( \frac{\partial \mathbf{w}}{\partial \mathbf{z}}%
\right) P_{D}(\mathbf{z})\left( \frac{\partial \mathbf{w}}{\partial \mathbf{z%
}}\right) ^{T}
\end{equation*}%
with $T$ denoting the matrix transpose. We find the nonzero commutators%
\begin{equation}
\left\{ x,p_{x}^{0}\right\} _{f}=\left\{ y,p_{y}^{0}\right\} _{f}=1\text{, \
\ \ }\left\{ \dot{x},\dot{y}\right\} _{f}=\frac{1}{\lambda }  \label{candir}
\end{equation}%
in coordinates $(x,y,\dot{x};p_{x}^{0},p_{y}^{0},\dot{y})$ adapted for $%
T^{\ast }D_{f}$.

\subsection{Reconstruction}

We first describe solutions geometrically and then give results for
analytical solutions for this is an easy linear system. The characteristics
of Eq.(\ref{lpcan}) are the Hamiltonian function (\ref{hdred}) and the
Casimir function%
\begin{equation*}
(J_{X}^{D}-\frac{l^{2}}{m})^{2}+(J_{Y}^{D})^{2}=R_{C}^{2}
\end{equation*}%
whose level sets are concentric cylinders of radius $R_{C}^{2}$ extending
along $J_{R}^{D}$ axis. Indeed, the solution curves are helices on the
circular paraboloids of revolution (\ref{paraboloid}), that is, orbits of
coadjoint action, which can be parametrized by%
\begin{equation*}
J_{R}^{D}=-\frac{\lambda }{2l^{2}}r^{2}\text{, \ \ }J_{X}^{D}=r\cos \theta 
\text{, \ \ }J_{Y}^{D}=r\sin \theta .
\end{equation*}%
The reduced Hamiltonian $H_{red}^{C}$ when further restricted to coadjoint
orbits, becomes 
\begin{equation*}
H_{\mathcal{O}}^{C}=-\frac{m}{2l^{2}}%
((J_{X}^{D})^{2}+(J_{Y}^{D})^{2})+J_{X}^{D}
\end{equation*}%
and the Hamilton's equations on coadjoint orbits%
\begin{equation*}
i_{X_{\mathcal{O}}}\omega _{\mathcal{O}}=dH_{\mathcal{O}}^{D}\text{, \ \ \ }%
X_{\mathcal{O}}=\dot{J}_{X}^{D}\frac{\partial }{\partial J_{X}^{D}}+\dot{J}%
_{Y}^{D}\frac{\partial }{\partial J_{Y}^{D}}
\end{equation*}%
consist of second and third equations in Eq.(\ref{lpcan}). To obtain
analytical solutions of these two equations, we get, by differentiation, 
\begin{equation*}
\text{\ }\ddot{J}_{Y}^{D}+\frac{m^{2}}{\lambda ^{2}}J_{Y}^{D}=0
\end{equation*}%
whose solution is%
\begin{equation*}
J_{Y}^{D}=A\sin \left( \frac{m}{\lambda }t\right) +B\cos \left( \frac{m}{%
\lambda }t\right) .
\end{equation*}%
Then, reduced equations imply the following solutions for the lifted
generators%
\begin{eqnarray*}
J_{X}^{D} &=&\frac{\lambda }{m}\dot{J}_{Y}^{D}+\frac{1}{m}l^{2}=-B\sin
\left( \frac{m}{\lambda }t\right) +A\cos \left( \frac{m}{\lambda }t\right) +%
\frac{l^{2}}{m} \\
J_{R}^{D} &=&\int J_{Y}^{D}(t)dt=B\frac{\lambda }{m}\sin \left( \frac{m}{%
\lambda }t\right) -A\frac{\lambda }{m}\cos \left( \frac{m}{\lambda }t\right)
-\frac{\lambda }{2l^{2}}(A^{2}+B^{2})-\frac{\lambda l^{2}}{2m^{2}}
\end{eqnarray*}%
where value of integration constant follows from the defining relation (\ref%
{jrd}). The velocities for base curves can be solved algebraically from the
definitions of orbit variables $J_{X}^{D}$ and $J_{Y}^{D}$%
\begin{eqnarray*}
\dot{x} &=&\frac{1}{l^{2}}(p_{x}^{0}J_{X}^{D}-p_{y}^{0}J_{Y}^{D})=A_{0}\cos
\left( \frac{m}{\lambda }t\right) -B_{0}\sin \left( \frac{m}{\lambda }%
t\right) +\frac{p_{x}^{0}}{m} \\
\dot{y} &=&\frac{1}{l^{2}}(p_{y}^{0}J_{X}^{D}+p_{x}^{0}J_{Y}^{D})=B_{0}\cos
\left( \frac{m}{\lambda }t\right) +A_{0}\sin \left( \frac{m}{\lambda }%
t\right) +\frac{p_{y}^{0}}{m}
\end{eqnarray*}%
with $A_{0}=\lambda (Ap_{x}^{0}-Bp_{y}^{0})/ml^{2}$ and $B_{0}=\lambda
(Bp_{x}^{0}+Ap_{y}^{0})/ml^{2}$. These first order reduced equations can now
be integrated to find solution curves on the configuration space $M$ 
\begin{eqnarray*}
x(t) &=&A_{0}\sin \left( \frac{m}{\lambda }t\right) +B_{0}\cos \left( \frac{m%
}{\lambda }t\right) +\frac{p_{x}^{0}}{m}t+C_{0x} \\
y(t) &=&B_{0}\sin \left( \frac{m}{\lambda }t\right) -A_{0}\cos \left( \frac{m%
}{\lambda }t\right) +\frac{p_{y}^{0}}{m}t+C_{0y}.
\end{eqnarray*}

\section{Discussions and conclusions}

Motivated by the equivalence proof of Marle \cite{marle} on Euler-Poincar%
\'{e} and Hamilton-Poincar\'{e} equations, we apply Hamiltonian reduction to
chiral oscillator described by a second order and degenerate Lagrangian \cite%
{luk},\cite{luk99}. Reduction with the appearent symmetry $\mathfrak{se}(2)$
of Lagrangian requires a central extension of $\mathfrak{se}(2)$ to
oscillator algebra $\mathfrak{osc}$ by Casimir function (\ref{cas}). This
provides a complete reduction of second order Lagrangian on a two
dimensional configuration space by a four dimensional Lie algebra of
symmetries.

The fact that the chiral oscillator exhibits extension of its symmetry
algebra was somehow known to the creators Lukierski, Stichel and Zakrzewski 
\cite{luk},\cite{luk99} of the model, because they designed the model in way
to have it \cite{hor02}. We showed necessity of central extension within the
geometric framework of Hamiltonian reduction theory.

The results of present work show that the chiral oscillator constitute an
example of a system as Poincar\'{e} dreamed it for first order Lagrangians 
\cite{poi01},\cite{marle}: the Euler-Lagrange equations can be reduced to a
first order system in generators of motion which, in turn, can be integrated
to obtain solution curves on configuration manifold.

More generally, given a second order Lagrangian on $T^{2}M$, the
Euler-Lagrange equations are equivalent to Hamilton's equations on $T^{\ast
}TM$ which are first order equations represented by a Hamiltonian vector
field $X_{H}$. The reduction procedure we employed identifies invariant
functions $J_{X_{a}}$, $X_{a}\in \mathfrak{g}$, such that $%
H=\sum_{a}b_{a}(t)J_{X_{a}}$ or, equivalently, $X_{H}=%
\sum_{a}b_{a}(t)F_{J_{X_{a}}}$. Moreover, the Poisson bracket algebra of
functions $J_{X_{a}}$ closes to form Lie-Poisson bracket algebra or,
equivalently, the Lie bracket algebra of lifted generators $F_{J_{X_{a}}}$.

Recently, Cari\~{n}ena, De Lucas, and Sard\'{o}n \cite{carinena13} studied
theory of Lie-Hamilton systems that seems to be related to the present work
in particular, and in general to reductions of Hamiltonian systems. It is
worth to mention their corresponding terminologies for further elaborations
of these techniques. $X_{H}$, being written as a linear combinations of $%
F_{J_{X_{a}}}$, is said to admit a superposition rule. In this case, $X_{H}$
is called a Lie system. The Lie-Scheffers theorem associates Lie systems to
finite dimensional Lie algebra of vector fields. The Lie algebra spanned by
the vector fields $F_{J_{X_{a}}}$ is called the Vessiot-Guldberg algebra.
The integration of such a Lie system as described in \cite{carinena13} is
precisely the reconstruction procedure for solution of Euler-Lagrange
equations explained above. The novelty of the cited work \cite{carinena13}
is that, their construction includes Lie algebra extensions in connection
with Casimir elements, a situation that emerged naturally in $SE(2)$
invariant chiral oscillator.

Lie-Poisson structure on dual of oscillator algebra was first obtained in 
\cite{vkm10} by reduction with $SE(2)$ of canonical symplectic structure
with magnetic term. In the present reduction of chiral oscillator
Lagrangian, extention to oscillator algebra follows also from reduction of
canonical structure but without magnetic term. Cendra, Marsden and Ratiu
mentioned in \cite{cmr04} possibility to view two-cocycle as magnetic term
arising by reduction and this view better explains the chiral oscillator
situation.

\section{Acknowledgements}

I am indepted to O\u{g}ul Esen and Serkan S\"{u}tl\"{u} for many
discussions. OE brougth the reference \cite{carinena13} to my attension. I
thank Professor \v{C}estm\'{\i}r Burd\'{\i}k, organizer of XII. Conference
on Quantum Theory and Symmetries in Prague where I had opportunity to
present early results leading the present work \cite{hgprag}.

\bigskip

\end{document}